\documentclass[runningheads]{llncs}

\usepackage[T1]{fontenc}
\usepackage{graphicx}
\usepackage{booktabs}
\begin{document}
\title{KV Cache Compression for Inference Efficiency in LLMs: A Review}
%
%\titlerunning{Yanyu Liu et al.}
\author{}
\institute{}
\author{Yanyu Liu\inst{1}\orcidID{0009-0000-1692-920X} \and
Jingying Fu\inst{1}\orcidID{0009-0003-9215-8250} \and
Sixiang Liu\inst{1}\orcidID{0009-0005-0149-991X} \and  
Yitian Zou\inst{1}\orcidID{0009-0008-4856-9489} \and
You Fu\inst{1}\orcidID{0000-0002-7809-4233} \and
Jiehan Zhou*\inst{1}\orcidID{0000-0002-4026-1649} \and
Shouhua Zhang*\inst{2}\orcidID{0000-0002-7980-5519}
}
\authorrunning{Yanyu Liu et al.}
\institute{Shandong University of Science and Technology, Qingdao 266590, China \\
\email{\{yanyu.liu,jingying.fu,sixiang.liu,yitian.zou,fuyou,jiehan.zhou\}@sdust.edu.cn}\\
\url{https://cogtwins.github.io/} \\
\and
University of Oulu, Oulu 90014, Finland \\
\email{shouhua.zhang@oulu.fi}\\
\**corresponding author}
\maketitle    % typeset the header of the contribution
\begin{abstract}
With the rapid advancement of large language models (LLMs), the context length for inference has been continuously increasing, leading to an exponential growth in the demand for Key-Value (KV) caching. This has resulted in a significant memory bottleneck, limiting the inference efficiency and scalability of the models. Therefore, optimizing the KV cache during inference is crucial for enhancing performance and efficiency. This review systematically examines current KV cache optimization techniques, including compression strategies such as selective token strategies, quantization, and attention compression. We evaluate the effectiveness, trade-offs, and application scenarios of these methods, providing a comprehensive analysis of their impact on memory usage and inference speed. We focus on identifying the limitations and challenges of existing methods, such as compatibility issues with different models and tasks. Additionally, this review highlights future research directions, including hybrid optimization techniques, adaptive dynamic strategies, and software-hardware co-design. These approaches aim to improve inference efficiency and promote the practical application of large language models. 

\end{abstract}
\section{Introduction}
%\subsection{A Subsection Sample}
In recent years, LLMs have revolutionized natural language processing by leveraging their ability to handle long contexts. This has enabled advancements in document summarization, detailed text analysis, multi-turn dialogue management, and complex code structure analysis, expanding their applications across various domains [1-3].

With the rapid expansion of LLMs context lengths, 128K-token contexts have become standard, and million-token contexts are now achievable [4]. As the context length increases, the demand for hardware resources, especially KV cache requirements, increases significantly. KV cache compression is a key technology for optimizing the inference efficiency of LLMs, primarily by compressing the key and value tensors in the self-attention mechanism to reduce memory usage and improve computational efficiency. Taking LLaMa as an example, the KV cache generated for each input token during decoding are reused by subsequent tokens to compute attention weights, avoiding redundant calculations. However, the memory usage of KV cache grows linearly with input sequence length and batch size, significantly increasing GPU memory requirements for long sequences or large-batch inference, becoming a performance bottleneck.

In order to solve the problem of excessive consumption of KV cache, in recent years, the academic and industrial have proposed a variety of compression methods. For example, some methods reduce the storage digit number of KV pairs through quantitative techniques [5], while others discard the unimportant tokens through the design of the clever elimination strategy [6]. Additionally, some methods reduce memory demand by removing unused KV entries from the cache of each attention head [7]. The emergence of these methods provides new ways to improve the inferencing efficiency of the LLMs. At present, two important reviews have systematically sorted out KV cache management technology. Shi et al. [8] and Li et al. [9] have reviewed the KV cache optimization technology of large language models from the perspective of stages and levels, respectively. The former focuses on the training, deployment and reasoning stages, and proposes relevant technical frameworks and evaluation indicators; the latter hierarchically expands the technical system and introduces multimodal evaluation. However, existing research is insufficient in cross-stage collaboration, dynamic adaptive strategies, and software and hardware co-design, making it difficult to meet the needs of efficient model deployment in complex scenarios.

The above shortcomings highlight the key gaps in current KV Cache optimization research in technology integration and scenario adaptability, and there is an urgent need to build a comprehensive methodology that covers the entire technology stack and takes into account static compression and dynamic regulation. The contributions of this article are as follows:

1. Present a comprehensive review on KV cache compression methods used in LLMs, including their principles, advantages, and limitations.

2. Analyze and compare their impact on performance and throughput in LLMs.

3. Suggest future research directions to enhance inference efficiency, including hybrid optimization techniques, adaptive dynamic strategies, and hardware-software co-design to optimize KV cache management and computational efficiency in LLMs.

The remainder of the paper is organized as follows: Section 2 presents an overview of existing KV cache compression techniques and analyzes their advantages and limitations. Section 3 compares their effects on model performance and throughput in LLMs. Section 4 and Section 5 makes a summary and highlights the future directions in intelligent strategies, technology integration, and engineering practice. 

%section{KV Cache Compression}
%subsection{KV selective compression}
\section{KV selective compressions}
Amidst rapid advancements in artificial intelligence, selective compression has emerged as a foundational optimization strategy for Large Language Models (LLMs). This methodology intelligently filters and compresses noncritical data, significantly reducing memory footprint while maintaining output fidelity. By prioritizing essential information, it effectively alleviates computational resource demands and enhances inference efficiency

Many cutting-edge achievements show strong advantages in this field. Zhong et al. proposed ZigZagKV [6], a dynamic KV cache compression method, which dynamically allocates budgets to compress the KV cache based on the uncertainty of layer attention and hidden state outputs. Yang et al. proposed KV Sharer [10], a layer-wise heterogeneous KV cache sharing strategy, which challenges conventional assumptions. Experiments demonstrated that KV Sharer reduces memory usage and improves inference speed while remaining compatible with other methods.Li et al. proposed EMS [11], a global-local scoring-based method for important token selection and adaptive compression, achieving state-of-the-art (SOTA) performance on LongBench and Needle-in-a-Haystack tasks.Yao et al. proposed CacheBlend [12], a cache fusion method for multi-text block inputs, which selectively recalculates the KV values of partial tokens to fuse the cache, reducing the first token generation time and increasing throughput while ensuring generation quality. Tang et al. proposed RazorAttention [13], a caching strategy based on attention head characteristics, which introduced compensation tokens to compress the KV cache without compromising performance.Chen et al. proposed NACL [14], a global optimization method for evicting tokens, combining proxy tokens and random eviction strategies to improve the performance of both long and short text tasks while reducing the KV cache.

Selective compression techniques exhibit multidimensional commonalities: critical token selection encompasses uncertainty-driven methods (ZigZagKV) and hybrid metric analysis (EMS); compression strategies incorporate retention, merging, and two-stage processes; while inter-layer relationship management involves differentiated approaches such as attention layer sharing (KV Sharer) and layer-wise budget allocation (ZigZagKV).Selective compression reduces memory consumption by choosing key KV cache, while maintaining the core functionality of the model. The aforementioned methods demonstrate superb skills in selecting key information and improve model performance. In order to visually present their performance, Table 1 presents a comparative analysis in terms of their throughput, inference efficiency, and compression ratio.

As shown in Table 1, representative methods achieve notable breakthroughs in KV cache optimization: memory footprint reduction exceeding 70\% (RazoratEntion [10]), inference throughput improvement of 2.8–5× (CacheBlend), and validated feasibility of multi-method collaborative optimization through the Needle-in-A-Haystack benchmark. Compared with conventional approaches, these methods exhibit systematic advantages in critical metrics including long-text processing, performance-memory trade-offs, and architectural adaptability. However, current single-dimensional optimizations based on cache selection still face bottlenecks in holistic performance enhancement.
\begin{table}[ht]
\caption{performance comparison among Selective Compressions}\label{tab1}
\centering
\scalebox{0.8}{ % 将表格缩小到原来的 80%
    \begin{tabular}{|c|c|c|c|c|}
        \hline
        Method & LLMs & \begin{tabular}[c]{@{}c@{}}Throughput\\ (times)\end{tabular} & \begin{tabular}[c]{@{}c@{}}Inference \\ efficiency(\%)\end{tabular} & \begin{tabular}[c]{@{}c@{}}Compression \\ ratio(\%)\end{tabular} \\
        \hline
        CacheBlend & \begin{tabular}[c]{@{}c@{}} LLaMa-70B\\ Mistral-7B\end{tabular} & 2.8 - 5 & 15 - 35 & N/A \\
        \hline % 添加此行，在CacheBlend与下一行之间添加横线
        RazorAttention & \begin{tabular}[c]{@{}c@{}} LLaMa-2\\  LLaMa-3\\ Qwen\end{tabular} & N/A & 10 & 70 \\
        \hline % 添加此行，在RazorAttention与下一行之间添加横线
        NACL & \begin{tabular}[c]{@{}c@{}} LLaMa 2-base\\  LLaMa2-Chat\end{tabular} & N/A & 80 & 50 \\
        \hline % 添加此行，在NACL与下一行之间添加横线
        KVShare & \begin{tabular}[c]{@{}c@{}} LLaMa2-7B\\  LLaMa2-13B\\ Mistral-7B\end{tabular} & N/A & 75 & 25 - 30 \\
        \hline % 添加此行，在KVShare与下一行之间添加横线
        LongBench & \begin{tabular}[c]{@{}c@{}} LLaMa -2-7B-Chat\\  LLaMa-3-8B-Instruct\\ Mistral-7B-Instruct-v0.2\end{tabular} & 6.74 & 28 - 79 & N/A \\
        \hline
    \end{tabular}
    }
\end{table}
%subsection{Quantitative compression}
\section{Quantitative compressions}
Quantization compression converts model keys and values from high-precision floating-point numbers to low-bit integers to reduce memory usage.This method effectively lowers memory and computational costs while maintaining inference accuracy,enabling efficient large-scale parallel computation and long-text processing.

There are many research in this field. Hooper et al. proposed KVQuant [15], which employs custom CUDA kernels for activation-aware quantization, boosting data throughput by 1.2× to 1.7× under low-precision conditions. Zhang et al. proposed Coupled Quantization (CQ) [16], a multi-channel joint quantization framework based on information-theoretic dependencies, which enhances encoding efficiency without compromising model stability.Liu et al. proposed KIVI [17], a 2-bit KV cache quantization algorithm that optimizes memory utilization, achieving 2.35× to 3.47× throughput gains with negligible performance loss. Tan et al. proposed AlignedKV [18], implementing precision-aligned adaptive quantization to minimize memory access overhead and accelerate attention computation with near-lossless accuracy.Dong et al. proposed QAQ [19], a sensitivity-guided non-uniform quantization method attaining 10× compression ratios across diverse tasks while maintaining model efficacy. Tao et al. proposed AsymKV [20], leveraging asymmetric and layer-wise quantization configurations to maximize memory efficiency for key-value matrices.Kim et al. proposed Lexico [5], approximating KV cache via sparse linear combinations with compact dictionaries, outperforming traditional methods in low-memory environments. These innovations collectively enhance storage-computation-energy efficiency while preserving core functionality, broadening LLMs applicability in real-world scenarios.

The above methods greatly improve the efficiency of memory utilization while maintaining the stability of model performance. To visually present their performance, Table 2 shows the comparative data of different quantization compression techniques in terms of throughput, perplexity, and compression ratio.
\begin{table}[ht]
\caption{performance comparison among Quantitative compressions}\label{tab2}
\centering % 添加此命令使表格居中
\scalebox{0.8}{ 
\begin{tabular}{|c|c|c|c|c|}
\hline
Method &
  LLMs &
  \begin{tabular}[c]{@{}c@{}}Throughput\\ (times)\end{tabular} &
  \begin{tabular}[c]{@{}c@{}}Perplexity\\ (\%)\end{tabular} &
  \begin{tabular}[c]{@{}c@{}}Compression \\ ratio(times)\end{tabular} \\ \hline
KVQuant    & \begin{tabular}[c]{@{}c@{}} LLaMa-7B\\  LLaMa-13B\\  LLaMa-65B\end{tabular}                        & 1.2 - 1.7   & 5.72 - 7.25   & 4.8   \\
\hline
KIVI      & \begin{tabular}[c]{@{}c@{}} LLaMa-2\\ Falcon\\ Mistral\end{tabular}                              & 2.35 - 3.47 & 12.74 - 63.05 & 2.6   \\
\hline
QAQ       & \begin{tabular}[c]{@{}c@{}} LLaMa2-7B\\  LLaMa2-13B\end{tabular}                                & N/A       & N/A         & 10    \\
\hline
AlignedKV & LLaMa-2-7B                                                                                      & N/A       & N/A         & 10    \\
\hline
LEXICO    & \begin{tabular}[c]{@{}c@{}} LLaMa3-8B\\  LLaMa-3-1B-Instruct\\  LLaMa-3-2-3B-Instruct\end{tabular} & N/A       & 48.29       & N/A   \\ \hline
AsymKV    & \begin{tabular}[c]{@{}c@{}} LLaMa-2-7B\\  LLaMa-2-13B\end{tabular}                                & N/A       & 58.12       & 6.7 - 8
\\ \hline

\end{tabular}
}
\end{table}
%subsection{Attention compression}
\section{Attention compressions}
Attention compression focuses on optimizing the KV cache in attention mechanisms to reduce memory usage and improve inference speed. Attention compression revises LLMs attention mechanisms, reducing KV cache memory usage by >40\% while retaining semantic precision. Unlike quantization-induced distortions or selective filtering limitations, this approach establishes robust computational topologies for ultra-large models.

Zhang et al. proposed H2O [21], dynamically modeling KV cache eviction as a submodular optimization problem to achieve >40\% throughput gains in LLaMa models, synergizing with quantization techniques. Adnan et al. proposed Keyformer [22], leveraging attention weight long-tail distributions to filter critical tokens, reducing GPT-J inference latency by 2.1× and boosting throughput by 2.4×. 
Zhao et al. proposed ALISA [23], combining hybrid sparse attention algorithms with INT8-quantized dynamic scheduling to triple LLaMa throughput versus FlexGen with <0.5\% accuracy degradation. Wang et al. proposed SQUEEZEATTENTION [24], compressing KV caches by 70\% via layer-wise cosine similarity analysis while limiting Mistral-7B performance variance to <1.2\%. Yang et al. proposed PyramidInfer [25], implementing hierarchical context compression under redundancy assumptions to expand LLaMa 2 batch sizes by 30\% and slash memory usage by 45\% in long-text tasks. Ma et al. proposed POD [26], clustering similar layers and optimizing attention aggregation to elevate LLaMa3-8B Haystack task accuracy by 8.7\% while reaching 93\% of theoretical batch limits. Jin et al. proposed L0-Ortho [27], fusing orthogonal transformations with distillation training to compress Llama 2-7B key-value heads by 87.5\%, recovering 99.3\% post-fine-tuning performance on BoolQ tasks.

These methods collectively advance KV cache optimization through memory reduction, accelerated inference, and performance preservation, providing critical technical foundations for scalable LLMs deployment.The above methods effectively reduce memory usage and accelerate inference speed. To visually present their performance, Table 3 shows the comparative data of different attention compression techniques in terms of throughput, inference efficiency, and compression ratio.

Table 3 demonstrates that attention compression have made certain progress in reducing memory occupation, improving the efficiency of reasoning, and maintaining model performance. However, there are still some challenges and limitation, such as the generality in different models and tasks, the compression efficiency, and the trade-off between compression and model performance.

\begin{table}[ht]
\caption{performance comparison among Attention compressions}\label{tab3}
\centering
\scalebox{0.8}{ % 将表格缩小到原来的 80%
\begin{tabular}{|c|c|c|c|c|}
\hline
Method &
  LLMs &
  \begin{tabular}[c]{@{}c@{}}Throughput\\ (times)\end{tabular} &
  \begin{tabular}[c]{@{}c@{}}Inference\\ efficiency(\%)\end{tabular} &
  \begin{tabular}[c]{@{}c@{}}Compression \\ ratio(times)\end{tabular} \\ \hline
H2O              & \begin{tabular}[c]{@{}c@{}}OPT-6.7B\\ OPT-30B\end{tabular}                                   & 2.3-3   & 18-73 & 5-10 \\  \hline
Keyformer        & \begin{tabular}[c]{@{}c@{}}GPT-NeoX-20B\\ GPT-J-6B\\ Cerebras-GPT-6.7B\\ MPT-7B\end{tabular} & 2.0-2.4 & 50-70 & 2.9  \\  \hline
ALISA            & \begin{tabular}[c]{@{}c@{}}OPT\\  LLaMa\end{tabular}                                          & 1.4-3.0 & N/A   & N/A  \\  \hline
SQUEEZEATTENTION & \begin{tabular}[c]{@{}c@{}}Mistral-7B\\  LLaMa2-70B\\ Mistral-7B\end{tabular}                 & 1.4-2.2 & 60    & N/A  \\   \hline
PyramidInfer     & \begin{tabular}[c]{@{}c@{}} LLaMa2-13B\\  LLaMa2-70B\end{tabular}                            & 1.7-2.8 & N/A   & N/A  \\ \hline
POD              & \begin{tabular}[c]{@{}c@{}} LLaMa3-8B-32K\\  LLaMa3.1-8B\end{tabular}                         & N/A     & N/A   & 1.54   \\  \hline
\end{tabular}
}
\end{table}
%subsection{Hybrid methods (based on LLaMA3-8B model)}
\section{Hybrid methods (based on LLaMA3-8B model)}
With the rapid increase in LLM scale, traditional single compression technology is facing bottlenecks. The advantages of hybrid compression, attention fusion, quantization, and other methods enable multidimensional optimization while maintaining performance. Main path: (1) Collaborative optimization of attention and data filtering to enhance long context reasoning ability; (2) Quantitative compression and computational optimization complement each other, efficiently managing memory; (3) Cross modal dynamic compression strategy enhances multimodal adaptability.

Lin et al. proposed DistAttention [28], which achieves dynamic computational allocation by decoupling resource scheduling between attention and non attention layers. Experiments have shown that processing 2000K tokens on 32 A100 GPU clusters can increase throughput by 1.35× to 3.4×.Kang et al. proposed the GEAR framework [29], which integrates quantization and error compensation techniques, and improves FP16 cache performance by 24.42\% under 2-bit quantization through low rank matrix and sparse matrix mechanisms. The experiment showed that its memory usage decreased by 2.39× and throughput increased by 2.1× to 5.07×.Wan et al. proposed a text-first compression strategy [30] for multimodal scenarios and designed a dynamic merging algorithm through modal attention difference analysis(LOOK-M). In the task, LOOK-M improves decoding speed by 1.5×, but the fine-grained control of cross modal feature interaction still needs improvement.Zhang et al. proposed Product Quantization (PQ) [31] technology to KV cache management and constructed a key-value retrieval system based on Maximum Internal Product Search (MIPS). PQCache combines effectiveness and efficiency. Even if only 1/5 of the labels participate in attention calculation, PQCache can maintain model quality while achieving acceptable system latency.

Existing methods improve cache utilization and throughput but face limitations: (1) hardware adaptation bottlenecks; (2) parameter sensitivity; (3) limited cross-modal generalization. Future work should focus on hybrid architectures with dynamic weight allocation, integrating resource scheduling and quantitative compression to develop adaptive systems as a key direction.

\section{Comparison}
In the application of the Llama model, different KV cache optimization demonstrate varying effectiveness in the improvement of model performance and throughput. This section presents a comparative analysis in terms of model performance and model throughput.
\subsection{Model performance}
Figure 1 illustrates the performance improvement of model inference rate achieved by optimization methods including KVSharer, NACL, RazorAttention, CQ, and KVQuant activation in the LLaMa model.

\begin{figure}
\includegraphics[width=0.68\textwidth]{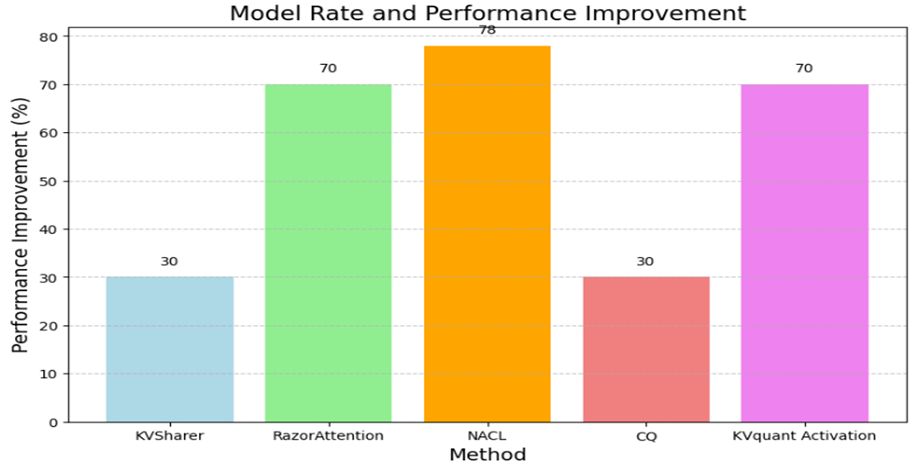}
\centering
\caption{The comparison in terms of model performance} \label{fig1}
\end{figure}
Figure 1 illustrates that the NACL method achieved the highest inference rate improvement (78\%) in the LLaMa model, highlighting its effectiveness in model optimization. In contrast, the KVSharer and CQ methods show relatively lower improvements (30\%). This analysis aids in selecting optimal methods based on specific deployment requirements, providing empirical support for enhancing LLMs inference performance.
\subsection{Model throughput}
In the LLMs inference, throughput serves as a key metric for evaluating computational efficiency, reflecting the system's capability to process textual inputs per unit time. Figure. 2 illustrates the throughput improvements achieved by optimization techniques including CacheBlend, Distributed Hybrid, and the KIVI Quantization in the LLaMa model.

\begin{figure}
\includegraphics[width=0.7\textwidth]{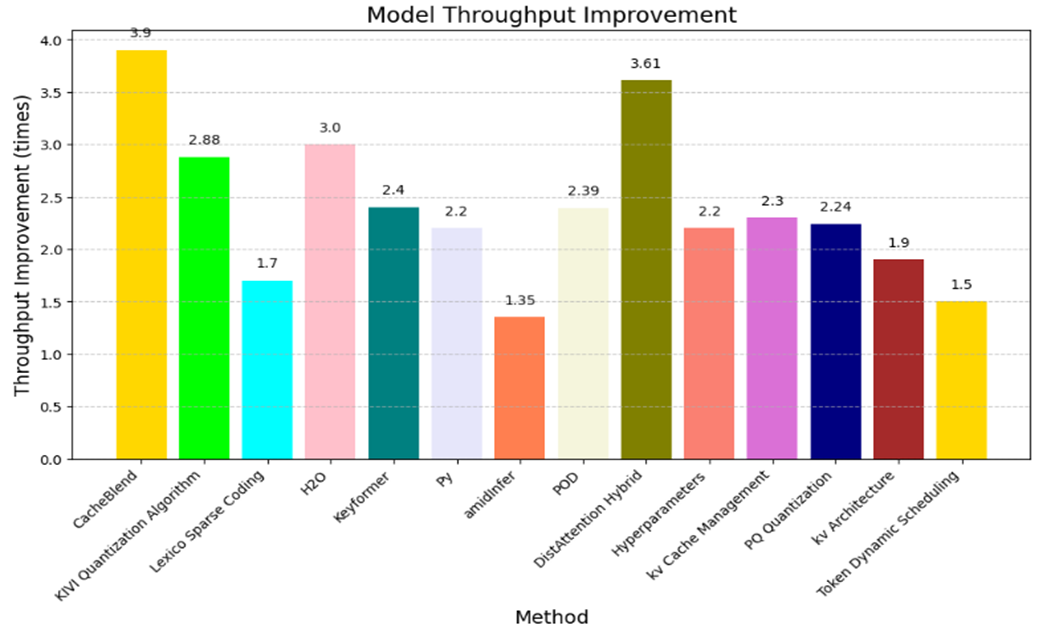}
\centering
\caption{The comparison in terms of model performance} \label{fig2}
\end{figure}
As shown in Figure 2, the CacheBlend and DistAttention hybrid methods achieve the most substantial throughput enhancements, demonstrating 3.9× and 3.61× improvements, respectively. In contrast, the Token Dynamic Scheduling method yields a comparatively modest 1.5× gain. These findings provide critical insights to identify optimal technical solutions that maximize throughput under specific operational constraints.

\section{Future Research Directions}
Based on the analysis and comparison of KV cache compression in LLaMa models, there are multiple optimization approaches. This paper focuses on three key development directions: Hybrid Optimization and Integration, Adaptive Dynamic Optimization, and Hardware-Software Co-Optimization. The following sections will explore each direction in terms of implementation strategies, challenges faced, and research priorities. 
\subsection{Hybrid optimization and integration}
This hybrid method enables optimized performance across various tasks and models, ensuring a balance between resource utilization and computational effectiveness. For example, the integration of selective compression, quantization compression, and attention compression allows for a more efficient caching strategy. By dynamically adjusting cache management, memory usage can be minimized while maintaining high inference efficiency.

Challenges: Compatibility issues arise when integrating different technologies. Quantization compression alters data accuracy, which may conflict with the attention compression's requirements for capturing key information. Additionally, the selective compression filtering strategy may not align with the low-precision data processing methods used after quantization. This makes it challenging to balance and coordinate these methods, ultimately limiting the overall performance improvement.

Research direction: A universal fusion framework are crucial for optimizing LLMs performance. This framework needs to intelligently adjust the priority, weight, and sequence of different compression techniques based on model architecture, task requirements, and computational constraints. By leveraging the strengths of each method while minimizing compatibility conflicts, it enhances overall optimization efficiency and ensures robust, adaptive performance across diverse applications.
\subsection{Adaptive dynamic optimization}
Research on dynamically adjusting cache compression is another promising direction. By tailoring compression strategies in real-time based on workload characteristics, task complexity, and model requirements, this approach can optimize memory efficiency, reduce latency, and enhance overall inference performance, ensuring adaptive and efficient cache management across diverse applications.

Challenges: Accurately predicting task requirements is difficult. Natural language processing tasks are complex and dynamic, making it challenging to predict accurately based on existing data and models. Additionally, dynamic adjustments generate additional resource overhead, and frequent adjustments can easily consume excessive computing resources and time, negatively impacting the optimization effect.

Research direction: Leverage machine learning and deep learning to analyze historical task data, input characteristics, and other relevant factors to develop a high-precision task demand prediction model for more efficient resource allocation and optimization. This enables the proactive and rational adjustment of caching strategies. Additionally, it is important to optimize resource management to minimize overhead while ensuring sustained or improved performance and efficiency.
\subsection{Collaborative optimization of hardware and software}
This collaboration is another research path. By integrating hardware-level support with KV cache optimization techniques, leveraging TPU/GPU acceleration, and applying KV cache compression methods, memory bottlenecks can be alleviated, leading to enhanced computational efficiency and overall performance.

Challenges: Adapting hardware and software is difficult. Different hardware architectures and features vary significantly, making it challenging to generalize software optimization solutions. Additionally, hardware upgrades occur rapidly, while, delayed software updates hinder the timely and effective utilization of hardware acceleration benefits.

Research direction: The promotion of collaborative hardware and software design can be achieved by establishing unified interface standards and collaboration specifications, allowing software to better adapt to hardware characteristics for enhanced optimization. Additionally, enhancing coordination between hardware and software R\&D ensures that hardware upgrades are synchronized with software optimization, maximizing performance efficiency.
\section{Conclusion}
This paper comprehensively reviews the KV cache compression techniques in LLMs. Given that the increase of LLMs context length leads to a sharp increase in the demand for KV cache resources, optimizing KV cache become extremely critical. Selective compression enables to screen important information to reduce memory usage and improve long text processing performance. Quantization compression converts key value accuracy to reduce costs without affecting complex task processing. Attention compression analyzes attention-related data compression cache to improve reasoning efficiency. Hybrid methods combine their advantages but also faces limitations. Through comparison with the LLaMa model, this review provides insights for selecting methods while analyzing and comparing the impact of various methods on model performance and throughput. In addition, this paper outlooks that, KV cache compression  should evolve towards hybrid optimization, adaptive dynamic optimization, and hardware-software co-optimization. Additionally, emphasis should be placed on intelligent co-design strategies, guiding future research in both theoretical exploration and practical applications. 

%\begin{credits}
%\subsubsection{\ackname} This work was support in part by the China NSFC under Grant 62072287; in part by the China NSFC under Grant W2412090; in part by the ghFund under Grant 202407027775; in part by the project ZR2024ME230 supported by Shandong Provincial Natural Science Foundation.

%\subsubsection{\discintname}
%The authors have no competing interests to declare that are relevant to the content of this article.

%\end{credits}

\end{document}